\pdfoutput=1 
\documentclass[aps,pra, letterpaper, notitlepage, floatfix, reprint, longbibliography]{revtex4-1}

\usepackage{amsmath,amsfonts,amssymb}

\usepackage{float} 

\usepackage{graphicx}
\usepackage[subrefformat=parens, caption=false]{subfig}

\usepackage{hyperref}
\hypersetup{colorlinks=true}
\usepackage[all]{hypcap} 

\usepackage{times}

\usepackage{upgreek}

\newcommand{\figurepanel}[2]{\hyperref[#1]{\ref*{#1}(#2)}}

\allowdisplaybreaks[1] 

\begin{document}
\title{Photon Correlations Generated by Inelastic Scattering in a One-Dimensional Waveguide Coupled to Three-Level Systems}
\author{Yao-Lung L. Fang}
\author{Harold U. Baranger}
\affiliation{Department of Physics, Duke University, P.O.~Box 90305, Durham, North Carolina 27708-0305, USA}
\date{\today}

\begin{abstract}
We study photon correlations generated by scattering from three-level systems (3LS) in one dimension. The two systems studied are a 3LS in a semi-infinite waveguide (3LS plus a mirror) and two 3LS in an infinite waveguide (double 3LS). Our two-photon scattering approach naturally connects photon correlation effects with inelastically scattered photons; it corresponds to input-output theory in the weak-probe limit. At the resonance where electromagnetically induced transparency (EIT) occurs, we find that no photons are scattered inelastically and hence there are no induced correlations. Slightly away from EIT, the total inelastically scattered flux is large, being substantially enhanced due to the additional interference paths. This enhancement carries over to the two-photon correlation function, which exhibits non-classical behavior such as strong bunching with a very long time-scale.
The long time scale originates from the slow-light effect associated with  EIT. 
\end{abstract}


\maketitle

\section{Introduction}

The many similarities between quantum transport of electrons (conduction) and optical phenomenon (propagation of EM radiation) have been used over the years to enrich both fields. While a scattering approach to the propagation of light, with input and output amplitudes, is quite natural in both classical and quantum optics \cite{JacksonEM, WallsMilburnQO08}, a comparable approach to electronic phenomena developed slowly. First introduced by Landauer \cite{LandauerIBM57, LandauerZPhys87}, it was subsequently substantially developed by B{\"u}ttiker \cite{ButtikerPRL86, ButtikerPRB88, ButtikerPRL90}. This approach was then used, for instance, to develop parallels in mesoscopic physics between electronic and photonic phenomena, such as coherent backscattering of electrons or photons from disordered media \cite{Akkermans-Montambaux, PingSheng}. Another example is in the development of semiclassical (or eikonal) approximations to quantum chaotic phenomena and the inclusion of diffractive effects \cite{Brack-Bhaduri}. While these parallels were developed mainly in the non-interacting-particle or linear-optics regime, interacting particles and the corresponding nonlinear regime are, of course, of key interest in both photonic and electronic transport. One particular setting that has received a great deal of attention in the quantum transport community is one-dimensional (1D) electrons interacting with local quantum impurities, a setting that includes for instance the Kondo problem, Anderson impurity model, and Bethe-Ansatz solutions \cite{Bruus-Flensberg, HewsonBook, GiamarchiBook}. The parallel photonic system is a one-dimensional EM waveguide strongly coupled to discrete non-linear quantum elements such as atoms, quantum dots, or qubits;  in analogy with ``cavity QED'' \cite{WallsMilburnQO08, HarocheRaimondBook}, the study of such systems is known as ``waveguide QED.'' 
 
The study of waveguide QED has increased rapidly over the past decade. Prior to that, there were a few early papers on the subject \cite{YudsonJETP84, YudsonJETP85, LeClairPRA97, KojimaPRA03, CamaletEPL04} that, for instance, exploited many-body approaches developed for electronic problems. The dramatic increase in interest starting in the period 2005-2008 \cite{ShenPRL05, ShenPRL07, ChangNatPhy07, KoshinoPRL07, YudsonPRA08} was driven by experimental progress toward achieving strong coupling between the waveguide and the local quantum system. Indeed, several experimental waveguide-QED platforms are being actively pursued. These include superconducting qubits coupled to a microwave transmission line \cite{AstafievSci10, AbdumalikovPRL10, HoiPRL12, vanLooScience13, InomataPRL14, HoiNatPhy15}, semiconductor quantum dots coupled to either a metallic nanostructure \cite{VersteeghNatCommun14, AkselrodNatPho14} or a photonic-crystal waveguide \cite{LodahlRMP15}, and more traditional quantum optics settings in which atoms provide the local quantum system and the waveguide is an optical fiber or glass capillary \cite{ChangVuleticLukinNP14, FaezPRL14}. Interesting waveguide-QED effects occur when the coupling to the waveguide dominates other emission or dephasing processes. Experiments in this interesting regime have been performed in several of the above waveguide-QED platforms. 

Two aspects of waveguide QED have attracted particular attention theoretically: the manipulation of single photons and the production of non-classical light. In the single photon arena, a variety of devices have been proposed that build on the manipulation of single photons by qubits or three-level systems (3LS) that is possible in 1D systems; for representative work in this area see Refs.~\cite{ChangVuleticLukinNP14, LodahlRMP15, HoiNJP13} and references therein. With regard to non-classical light, the main characteristics studied are the photon-photon correlation function (also called the second-order coherence \cite{LoudonQTL03}) and the photon statistics. The majority of work on these topics has treated a single quantum system coupled to the waveguide, where the single quantum system is modeled as a two-level system (2LS) or the only slightly more complicated driven 3LS (for very recent work along these lines see, for example, Refs.\,\cite{RoyPRA14, LiSunPRA14, PletyukhovPRA15, XuFanPRA15}). Correlation effects in a multi-qubit waveguide have been studied in a number of recent papers using a variety of techniques \cite{DzsotjanPRB10, DzsotjanPRB11, TudelaPRL11, CanoPRB11, ChangNJP12, LalumierePRA13,CanevaChangarXiv15, ZhengPRL13, FangEPJQT14, FangPRA15, LaaksoPRL14, ZuecoFD14,ShiChangCiracX15, GuimondarXiv15}. In most of these, the Markovian approximation is required in order to simplify the interactions between the qubits via the waveguide \cite{DzsotjanPRB10, DzsotjanPRB11, TudelaPRL11, CanoPRB11, ChangNJP12, LalumierePRA13, CanevaChangarXiv15}. There are, however, a few non-Markovian results \cite{ZhengPRL13, FangEPJQT14, FangPRA15, LaaksoPRL14, ZuecoFD14, ShiChangCiracX15, GuimondarXiv15} which have been used to delineate the range of validity of the Markov approximation. 

Here we extend our recent results on multiple 2LS waveguide QED \cite{ZhengPRL13, FangEPJQT14, FangPRA15} to the case in which driven 3LS are used. We calculate the two-photon wavefunction and focus on photon-photon correlations. We find that these correlations are substantially enhanced in systems containing 3LS, making them better experimental candidates for further study of the non-classical light produced. Furthermore, we find that the complexity of the structure enhances the photon-photon correlations---they are enhanced by adding additional nonlinear elements (qubits) as well as by simply adding a mirror. The photons can be either bunched or anti-bunched depending on the situation, and we find cases of both strong bunching and anti-bunching. 

The paper is organized as follows. In the next section, we first recap the standard model of waveguide QED and a 3LS and summarize our approach to finding the two-photon wavefunction. Then we present the physical quantities that are calculated, emphasizing the total inelastic scattering as a measure of the correlated part of the wavefunction. Results for a single 3LS are presented in Section 3 as a basis for comparison to the more complex structures studied later. In Section 4 we add a mirror to the system, thus studying a single 3LS in a semi-infinite waveguide. Section 5 covers results for two 3LS in an infinite waveguide. In the results of both Sections 4 and 5, inelastic scattering is enhanced, suggesting more visible correlation effects. Finally, in Section 6 
we discuss implications of the results and conclude.

\section{Model and Observables}

\subsection{Waveguide QED model with multiple three-level systems}

\begin{figure}
	\centering
	\includegraphics[width=7cm]{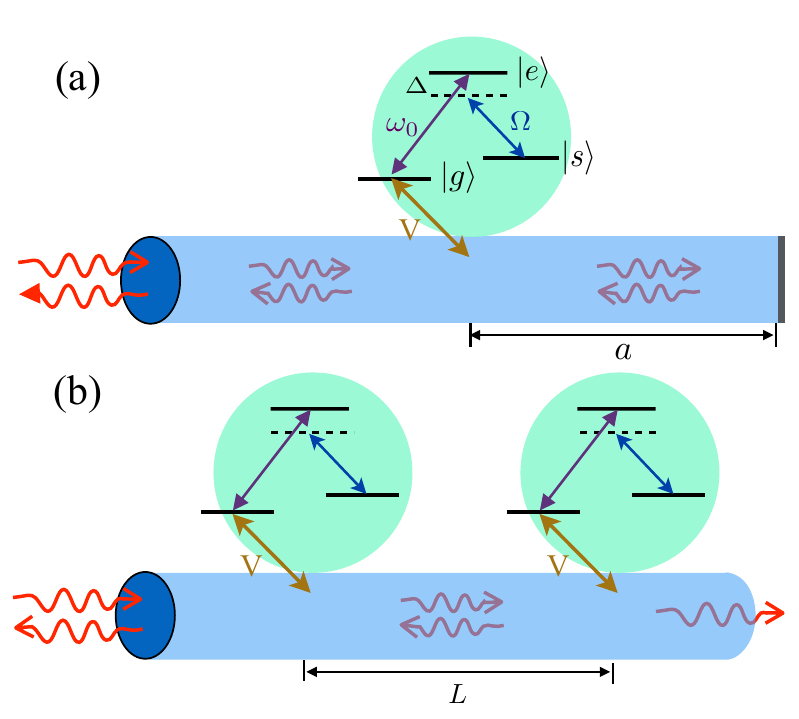}
	\caption{\textbf{Schematic of waveguide QED.} (a) Single 3LS coupled to a semi-infinite waveguide with qubit-mirror separation $a$, $\omega_e=\omega_0$ and $\omega_s=\omega_0-\Delta$; (b) Two identical 3LS, separated by distance $L$, coupled to an infinite waveguide.
	}
	\label{fig:schematic}
\end{figure}

The standard Hamiltonian of waveguide QED \cite{ShenPRL05, ChangNatPhy07} consists of a one-dimensional bosonic field that can travel to the left or right coupled to $N$ local quantum systems, often called simply qubits. For a schematic see Fig.~\ref{fig:schematic}. Within the rotating-wave approximation, the Hamiltonian in real space reads (taking $\hbar=c=1$)
\begin{align}
	&H=H_\text{QS}
	-i\int\limits_{-\infty}^{\infty} dx \left[a^\dagger_\text{R}(x)\frac{d}{dx}a_\text{R}(x)-a^\dagger_\text{L}(x)\frac{d}{dx}a_\text{L}(x)\right]\nonumber\\
	&+\sum_{i=1}^N\sum_{\alpha=\text{L,R}} \!V_i\! \int\limits_{-\infty}^{\infty} \!dx\; \delta(x-x_i)\left[a^\dagger_\alpha(x)\sigma^{(i)}_{ge}+\sigma^{(i)}_{eg} a_\alpha(x)\right],\label{eq:Hamiltonian}
\end{align}
where $\sigma^{(i)}_{eg}=|e\rangle_i\langle g|$ denotes the Pauli raising operator of the \textit{i}-th qubit with position $x_i$ and coupling strength $V_i$, and $a_\text{R,L}$ denote the annihilation operators of right- or left- going photons.  The corresponding decay rate of the \textit{i}-th qubit to the waveguide is $\Gamma_i \equiv 2V_i^2$. Throughout this paper, the coupling of all of the qubits is the same, $V$. In order to assess the maximum possible non-classical light effects that could be present, we focus on the lossless limit.

The local quantum systems that we consider here are identical 3LS,
$H_\text{QS} = \sum_i H_\text{3LS}^{(i)}$. 
The Hamiltonian for a $\Lambda$-type 3LS is 
\begin{equation}
H_\text{3LS}=\omega_0 |e\rangle\langle e|+\omega_s |s\rangle\langle s|+\frac{\Omega}{2} \left(|e\rangle\langle s|+|s\rangle\langle e|\right),
\end{equation} 
in which $\Omega$ is the Rabi frequency of the classical driving and 
$\omega_s=\omega_e-\Delta$ with $\Delta$ being the detuning between
the driving frequency and the frequency of the $|s\rangle$ to 
$|e\rangle$ transition. (The frequency corresponding to the ground state is taken to be zero.)
Finally, we note that a mirror can be introduced as a boundary condition when solving for the single-photon wavefunction \cite{PeropadrePRA11, KoshinoNJP12, TufarelliPRA13, FangPRA15}.  

To construct the two-photon scattering wavefunction, we use the Lippmann-Schwinger equation \cite{RoyPRA11, ZhengPRL13, FangEPJQT14, FangPRA15}, in which the Pauli raising and lowering operators, $\sigma_{eg}$ and $\sigma_{ge}$, are replaced by bosonic creation and annihilation operators, $b^\dagger$ and $b$. (A similar approach has been used in the case of two-electron scattering \cite{DharPRL08,RoyPRB08}.) To satisfy the level statistics, it is necessary to introduce an additional on-site repulsion $U$ to be taken as infinite at the end. For a 2LS, it is known that this approach correctly gives all measurable quantities \cite{ZhengPRL13, FangPRA15}. For a 3LS, in addition to repulsion for each upper level, an extra term has to be added so that the double occupancy can be fully ruled out: the repulsion operator $\tilde{V}$ is
\begin{equation}
\tilde{V}=\frac{U}{2}\left(b_e^\dagger b_e^\dagger b_e b_e + b_s^\dagger b_s^\dagger b_s b_s +2 b_e^\dagger b_e b_s^\dagger b_s \right).
\label{eq:on-site interaction}
\end{equation} 
Note that the coefficient of the last term is chosen for convenience; any coefficient would be canceled out after taking $U\rightarrow\infty$. Once a proper on-site interaction $\tilde{V}$ is introduced for each qubit, the calculation of the two-photon wavefunction $|\psi_2\rangle$ can be done straightforwardly \cite{FangEPJQT14, FangPRA15}. 

The two-photon scattering wavefunctions that emerge have a common structure (see, for instance, Ref.\,\cite{XuFanPRL13} and references therein): an incoming multi-particle plane wave and two contributions to the outgoing part--- (i) an outgoing multi-particle plane wave in which the momenta of the photons are just rearranged and (ii) a part that involves a continuum of momenta that cannot be written as a simple plane wave. The first part is elastic scattering, and the second is inelastic (for each individual particle). It is this latter part of the wavefunction that encodes the correlation between the two photons; it has been called the ``bound state part'' \cite{ShenPRL07, NishinoPRL09, ImamuraPRB09, ZhengPRA10, RoyPRA11} because as the distance between the particles grows, the wavefunction decays exponentially. In scattering theory, it corresponds to the two-particle irreducible T-matrix \cite{XuFanPRL13}. 

Having defined the model, we now turn to the observables that we study.

\subsection{Power spectrum (resonance fluorescence), $S(\omega)$}

The power spectrum is defined as the Fourier transform of the amplitude correlation function (also called the first-order coherence),
\begin{equation}
S_\alpha(\omega)=\int dt\,e^{-i\omega t}
\langle\psi_2| a_\alpha^\dagger(x_0) a_\alpha(x_0+t) |\psi_2\rangle,
\label{eq:def_S(w)}
\end{equation}
for $\alpha=\,$R or L, corresponding to right-going or left-going photons. $S_\alpha(\omega)$ gives simply the intensity of outgoing photons as a function of their frequency. The elastic scattering gives a $\delta$-function contribution which we drop; thus, in this paper $S_\alpha(\omega)$ refers to the \emph{inelastic} power spectrum.

\subsection{Total inelastically scattered component, $F(k_{\rm in})$}

Since in the two-photon wavefunction, the inelastically scattered component is necessarily the correlated ``bound state'' part of the wavefunction and vice versa, the total inelastically scattered power provides a measure of the overall strength of correlations in the situation studied. This, then, can be used as a figure of merit to compare different systems or optimize parameters in order to find the largest non-classical-light effects. Thus, we define a function of the incoming momenta,
\begin{equation}
F(k_\text{in,1},k_{{\rm in},2}) \equiv \int d\omega \,\left[S_\text{R}(\omega)+S_\text{L}(\omega)\right].
\label{eq:def_F(k)}
\end{equation}
In our study, we take the two incoming photons to have the same energy, and so we consider a function of one variable, $F(k_{\rm in})$. We shall refer to the maximum value of $F(k)$, denoted $F_\text{peak}$, and the value of $k$ at which this maximum is reached, $k_\text{peak}$.

\subsection{Photon-photon correlation (second-order coherence), $g_2(t)$}

A key signature of non-classical light is that the photons can be bunched or anti-bunched in space or time. This is revealed through the photon-photon correlation function, $g_2(t)$ \cite{LoudonQTL03}. Physically, $g_2(t)$ is proportional to the joint detection probability of two photons far from the scattering region that are separated by distance $ct$. It is customarily normalized by taking the ratio of the joint detection probability to the probability of measuring single photons independently; thus, $g_2(t)=1$ signifies no correlation. As the photons should be uncorrelated at very large separations in the systems we study, we have the property $g_2(t \rightarrow \infty)=1$ \cite{LoudonQTL03}.

For a weak coherent state, the correlation comes from the two-particle wavefunction while the normalization is given by the single-photon wavefunction (see, e.g., Refs.~\cite{FangEPJQT14,RoyPRA14}), yielding
\begin{equation}
g_2(t) = \frac{\langle\psi_2|a^\dagger_\alpha(x_0)a^\dagger_{\alpha}(x_0+t)a_{\alpha}(x_0+t)a_\alpha(x_0)|\psi_2\rangle}{|\langle\psi_1|a^\dagger_\alpha(x_0)a_\alpha(x_0)|\psi_1\rangle|^2}.
\label{eq:def_g2}
\end{equation}

We now turn to presenting results for these observables in three cases. 

\section{Single 3LS}

\begin{figure}[t]
	\centering
	\includegraphics{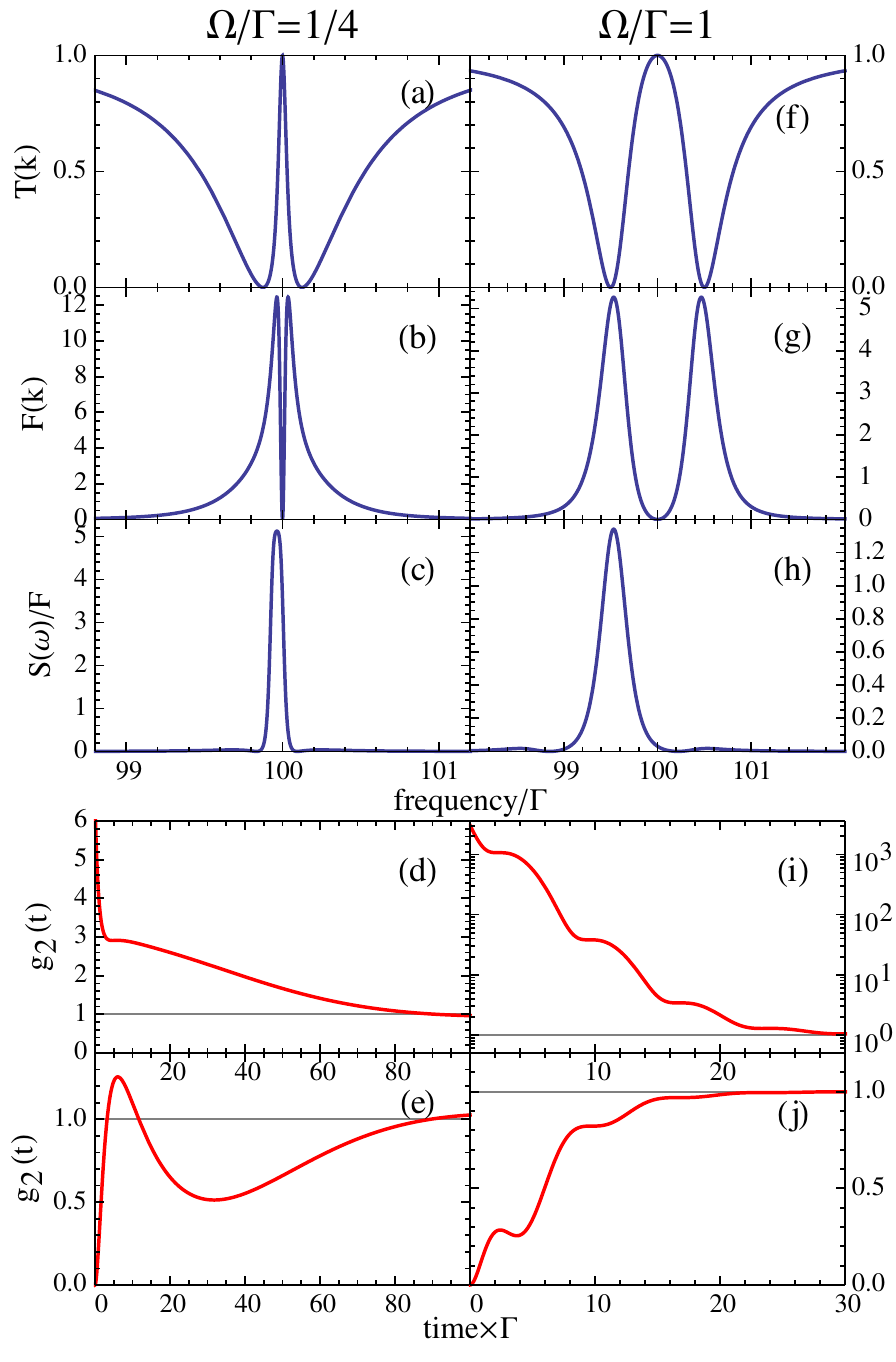}
	\caption{\textbf{Single 3LS coupled to an infinite waveguide:} single-photon transmission probability $T$, total inelastically scattered flux $F$, normalized power spectrum $S$, and correlation functions $g_2$ of the transmitted (fourth row) and reflected (fifth row) intensity. The classical Rabi frequency for the first column is $\Omega=\Gamma/4$ and for the second column is $\Gamma$. $T$ and $F$ are functions of the incident frequency $k$, while the power spectrum $S$ and the correlation function $g_2$ are calculated by fixing $k=k_\text{peak}$, the incident frequency giving rise to maximum $F$. Note that for a single 3LS, $S_\text{R}=S_\text{L}=S$. A horizontal line indicating the uncorrelated value $g_2=1$ is plotted. The transmittance at $k=k_\text{peak}$  for $\Omega/\Gamma=1/4$ is $T(k_\text{peak})=39.3\%$ and for $\Omega/\Gamma=1$ is $1.8\%$. (Parameters used: $\omega_0=\omega_s= 100\Gamma$ and $\Delta=0$.)}
	\label{fig:single 3LS}
\end{figure}

We start from a single 3LS coupled to an infinite waveguide, a system previously investigated in Refs.~\cite{TsoiPRA09, SillanpaaPRL09, WitthautNJP10, BianchettiPRL10, RoyPRL11, ZhengPRA12, LeungPRL12, MartensNJP13, KoshinoPRL13, RoyPRA14}. For two identical photons injected from the left, the results are shown in Fig.~\ref{fig:single 3LS}. 
When each photon is resonant with the 3LS, $k=\omega_s=\omega_e-\Delta$, electromagnetically induced transparency (EIT) occurs---an interference effect between different processes within the 3LS that leads to decoupling of the 3LS from the waveguide \cite{FleischhauerRMP05}. The width of the resulting peak in transmission is $\sim\Omega^2/\Gamma$ \cite{LeungPRL12, FleischhauerRMP05}. 
EIT is clearly seen in the single-photon transmitted intensity as a function of input wavevector $k$ in panels (a) and (f) of Fig.~\ref{fig:single 3LS}. 

As an illustrative case, we give the expressions for $S(\omega)$ and $F(k)$ in the Appendix, and plot them in panels (b), (c), (g), and (h). $F(k)$ has a zero at the EIT peak (when $k=\omega_s$) and two sharp peaks nearby. The fact that $F(k_{\rm in}=\omega_s)=0$ at perfect transparency is, in fact, true when any number of identical 3LS are coupled to the waveguide. This means that at this resonance, all of the photons are scattered elastically, and that, therefore, there is no ``bound state'' part of the wavefunction. This is consistent with the fact that there are no correlations in these cases---the correlation function of the transmitted photons is $g_2(t)=1$ \cite{ZhengPRA12, RoyPRA14}. This phenomenon has been called ``fluorescence quenching'' \cite{ZhouPRL96, RephaeliPRA11}.  

The surprisingly large peak value of $F$ in panel (b) suggests that there will be large correlation effects because of the necessary connection between inelastic processes and photon-photon correlation. For comparison, $F_\text{peak}$ for a single 2LS is $8/\pi\Gamma$: this can be derived either by setting $\Omega$ to zero in Eq.~\eqref{eq:inelastic flux single 3LS} or by multiplying Eq.~(D4) of Ref.~\cite{FangPRA15} by $4\pi$. Thus, when $\Omega/\Gamma=1/4$, $F_\text{peak}$ is about five times larger in the 3LS case than for a 2LS. We infer that a 3LS is much more effective than a 2LS in creating photon correlations. 


To explore the effect of the correlated part of the wavefunction, we consider the most favorable case: photons injected at the energy at which $F$ is maximized, $k=k_\text{peak}$. In the EIT regime (weak driving, $\Omega/\Gamma\ll1$), panel (c) shows that the power spectrum has a sharp peak centered at $k_\text{peak}$, and panels (d) and (e) show that the time scale of the correlation function is long, $\sim 40/\Gamma$. The slow decay time is associated with the time delay $\tau$ \cite{FleischhauerRMP05, FangPRA15}, which is itself associated with the inverse of the width of the EIT peak. At $k=\omega_s$, the time delay is simply $\tau=2\Gamma/\Omega^2$. 

An alternative route to obtaining a large $F_\text{peak}$ and strong correlations with a long decay time is to use multiple 2LS. From our previous work \cite{FangPRA15}, we note that in order to obtain the large value of $F_\text{peak}$ and slow decay obtained here for a single 3LS, one has to use of order ten 2LS---a considerably more complicated structure (but one that does not require classical driving). 

On the other hand, features typical for a single qubit, regardless of its level structure, coupled to an infinite waveguide are present. For example, the reflected $g_2(t=0)$ is always zero because a single atom cannot reflect two photons at the same time. Moreover, the transmitted and reflected power spectra are identical [for a single 3LS, see Eq.~\eqref{eq:power spectrum single 3LS}].

\begin{figure*}
	\centering
	\includegraphics{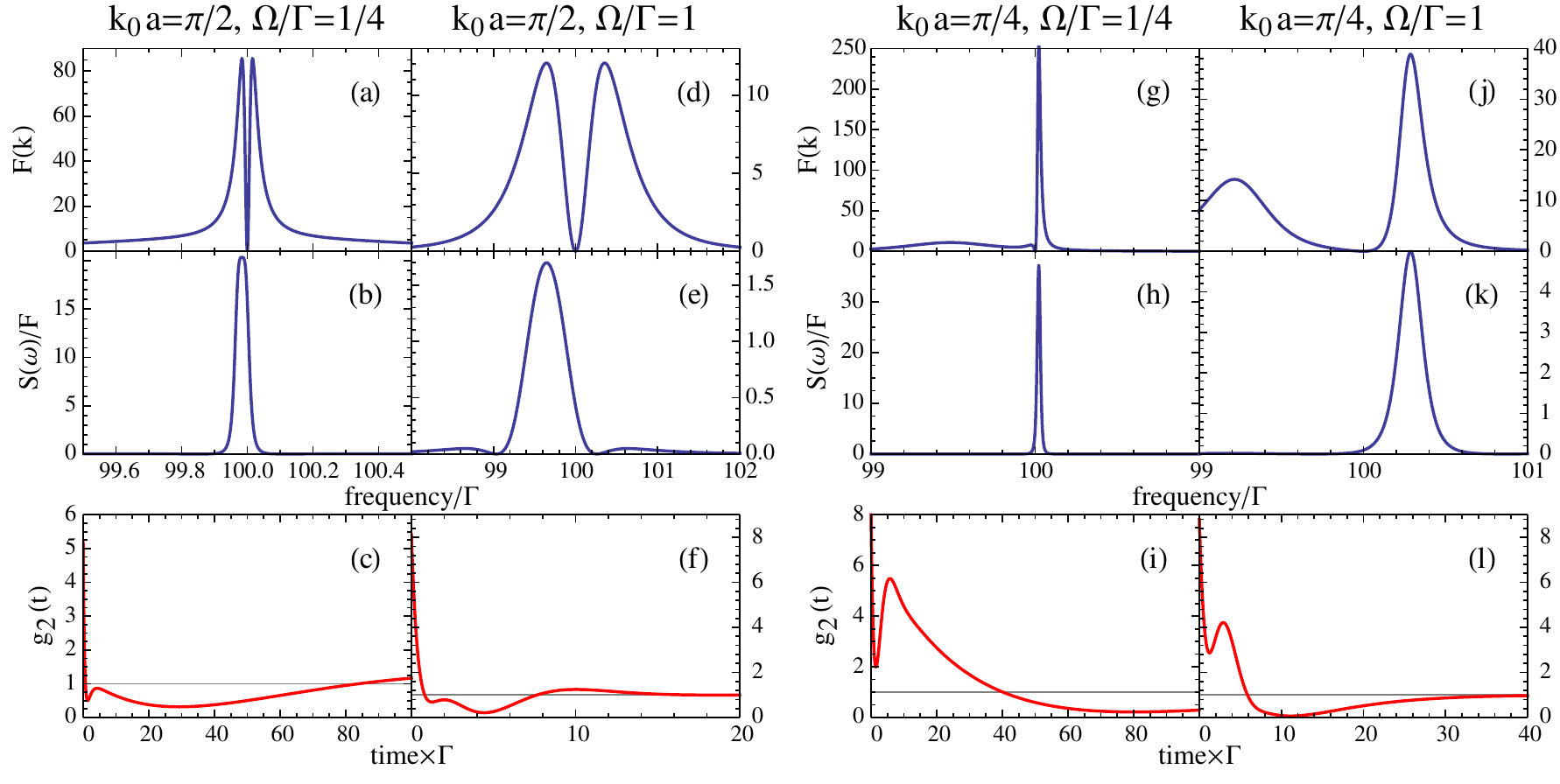}
	\caption{\textbf{Single 3LS coupled to a semi-infinite waveguide} [cf.\,Fig.\,\figurepanel{fig:schematic}{a}]: total inelastically scattered flux $F$, normalized power spectrum $S$, and reflected correlation function $g_2$. Qubit-mirror separation $a$ and classical Rabi frequency $\Omega$ used for each column are labeled on the top. $F$ is a function of the incident frequency $k$, while the power spectrum $S$ and the correlation function $g_2$ are calculated by fixing $k=k_\text{peak}$, the incident frequency giving rise to maximum $F$. A horizontal line indicating the uncorrelated value $g_2=1$ is plotted. (Parameters used: $\omega_0=\omega_s= 100\Gamma$ and $\Delta=0$.)}
	\label{fig:single 3LS semi-infinite}
\end{figure*}

The magnitude of the field driving the 3LS has a large effect on the correlation properties---compare the first and second columns of Fig.~\ref{fig:single 3LS}. First, the magnitude of $F$ clearly decreases, implying less correlation. In fact, for $\Omega \neq 0$, $F_{\rm peak}$ is a monotonically decreasing function of $\Omega$ [see Eq.~\eqref{eq:inelastic flux single 3LS}]. It can be very large in the EIT regime ($\Omega/\Gamma \ll 1$) and then decreases to $16/\pi\Gamma$ as $\Omega/\Gamma$ approaches one; curiously, $F_\text{peak}$ at $\Omega=0$---that is, in the 2LS case---is half this large $\Omega$ value.  Second, the spectrum of outgoing photons is broadened into a Lorentzian form. Finally, the timescale for decay of $g_2(t)$ is considerably shortened. The nature of the correlations at $\Omega/\Gamma \sim 1$ is such that there are few special features. Thus, using a 3LS with strong driving provides a route to relatively simple, less structured correlation effects, a feature that may be desirable in some situations. 

We close this section by commenting on the behavior of $F(k)$ at large $k$. Upon inspecting the expression for $F(k)$ in Eq.~\eqref{eq:inelastic flux single 3LS}, we find that it decays as $1/k^4$ for incident frequency $k$ far away from $\omega_s$. This form of decay is also true for a single 2LS ($\Omega=0$).

\section{3LS plus mirror}

To construct the second system that we study, we add a mirror to the right of the waveguide so that photons can enter or leave only from the left---see Fig.\,\figurepanel{fig:schematic}{a} for a schematic---thus, the system consists of a single 3LS coupled to a semi-infinite waveguide \cite{KoshinoNJP13}. As a result, the amplitude of the single-photon reflection is unity, $|r(k)|=1$. The calculation of the single- and two- photon eigenstates is explained in detail in Ref.\,\cite{FangPRA15} in the case of a 2LS; the 3LS case is, of course, very similar. Here we suppose the 3LS is placed close to the end of the waveguide, $k_0a\lesssim2\pi$ where $k_0=\omega_0/c=2\pi/\lambda_0$ is the wavevector associated with the qubit frequency $\omega_0$. We use, therefore, the Markovian approximation. 

Results are shown in Fig.~\ref{fig:single 3LS semi-infinite} for two values of the separation between the 3LS and the mirror: $k_0a=\pi/2$ and $\pi/4$ which correspond to $a=\lambda_0/4$ and $\lambda_0/8$. As for a single 2LS coupled to a semi-infinite waveguide (see, for example,
Refs.\,\cite{KoshinoNJP12, TufarelliPRA13, BradfordPRA13, HoiNatPhy15, FangPRA15} and references therein), interference effects associated with the qubit-mirror separation $a$ have a large effect on all measurable quantities. These effects can be connected to the complicated response of the poles of the reflection amplitude to the placement of the 3LS. For a discussion of the role of poles in the 2LS case see Ref.\,\cite{FangPRA15}; the present 3LS case is similar. 

We highlight several features of the results in 
Fig.~\ref{fig:single 3LS semi-infinite}:
\begin{enumerate}
\item The magnitude of inelastic scattering increases substantially upon adding a mirror: compare the first row of 
Fig.~\ref{fig:single 3LS semi-infinite} to the second row of Fig.~\ref{fig:single 3LS}. In the EIT regime, $F_\text{peak}$ is roughly an order of magnitude larger when a mirror is present. Note in addition that for both values of $a$, $F=0$ when the injected photons are on resonance with the 3LS. 

\item $F(k)$ is symmetric for $k_0a=\pi/2$ but is highly asymmetric for $k_0a=\pi/4$. The latter is connected to the asymmetric pole structure.

\item $S(\omega)$ is narrow and largely structureless, centered at and symmetric with respect to $k_\text{in}$ (see the second row in Fig.~\ref{fig:single 3LS semi-infinite}).

\item Strong initial bunching is found (third row), $g_2(t=0) \gtrsim 5$, in contrast to the initial anti-bunching found for reflection in an infinite waveguide---because a photon reflected by the mirror can stimulate emission from the 3LS, the reflected $g_2(0)$ need not be zero. The bunching is followed by anti-bunching characterized by a long time-scale. 

\item Increasing the classical driving field $\Omega$ causes a shift in $k_{\rm peak}$ away from the resonant frequency---which is natural as the EIT feature becomes broader---as well as a sharp decrease in the magnitude $F(k_{\rm peak})$. The magnitude of the secondary peak in the case $k_0a=\pi/4$ is less sensitive to $\Omega$ than the primary peak. The long time anti-bunching is cut-off for increasing $\Omega$. 

\end{enumerate}

One can derive the time delay $\tau$ for a semi-infinite waveguide by differentiating the phase $\phi_k$ of the reflection amplitude $r(k)$ with respect to $k$ \cite{deCarvalho02, FangPRA15}. Surprisingly, on resonance ($k=\omega_s$), the time delay has a very simple expression $\tau=2\times2\tilde{\Gamma}/\Omega^2$, where $\tilde{\Gamma}=\Gamma\left[1-\cos(2k_0a)\right]$ is the effective decay rate of the 2LS in front of the mirror \cite{KoshinoNJP12, FangPRA15, HoiNatPhy15}. This suggests that this system can be seen, roughly speaking, as a double 3LS (a real one and its mirror image) coupled to an infinite waveguide, a system to which we turn in the next section.

\section{Two 3LS}

\begin{figure*}
	\centering
	\includegraphics[width=7.0in]{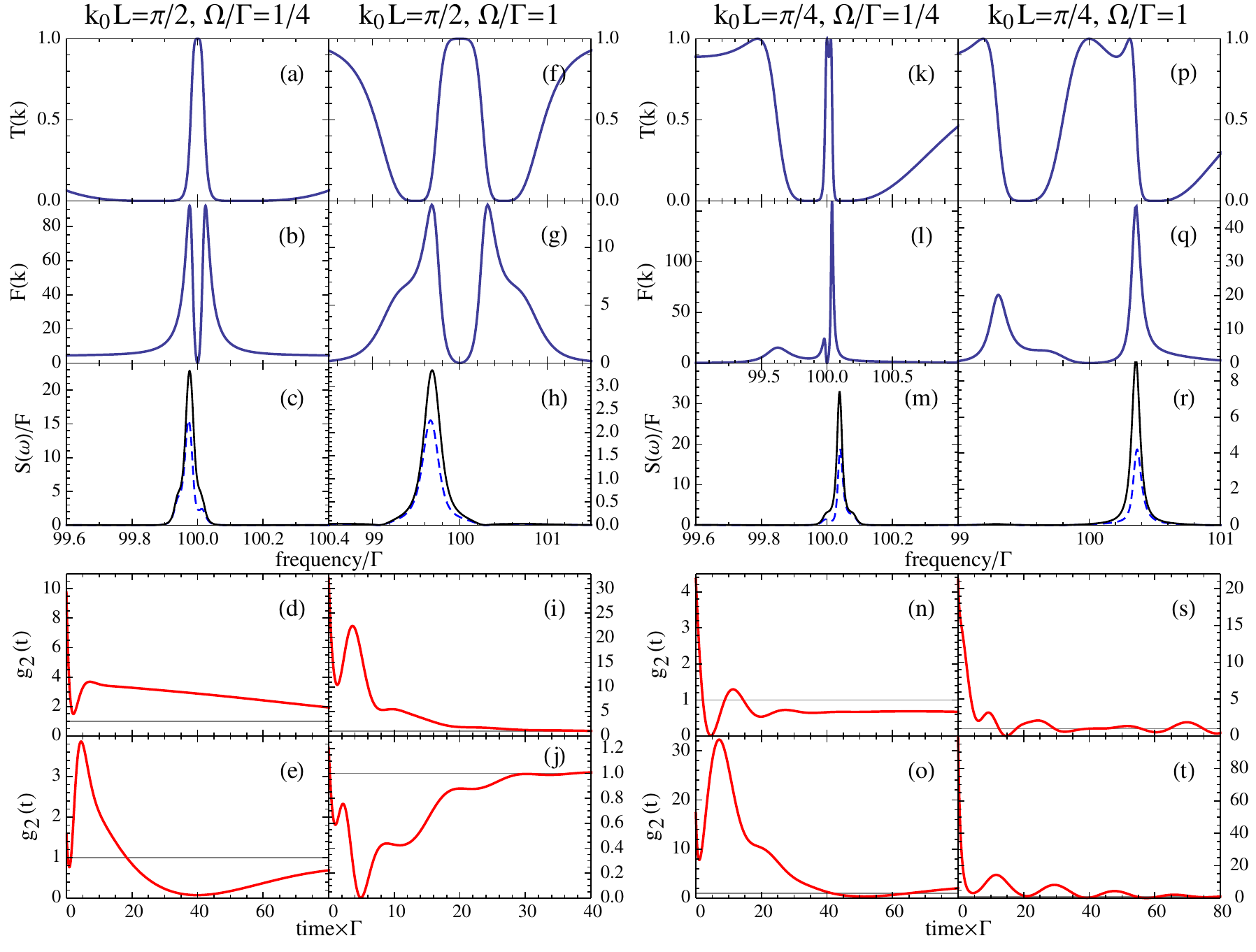}
	\caption{\textbf{Two 3LS coupled to an infinite waveguide} [cf.\,Fig.\,\figurepanel{fig:schematic}{b}]: single-photon transmission $T$, total inelastically scattered flux $F$, normalized power spectrum $S$, and the correlation functions $g_2$ for the transmitted (fourth row) and reflected (fifth row) intensity. Qubit-qubit separation $L$ and classical Rabi frequency $\Omega$ used for each column are labeled on the top. $T$ and $F$ are functions of the incident frequency $k$, while the power spectrum $S$ and the correlation function $g_2$ are calculated by fixing $k=k_\text{peak}$, the incident frequency giving rise to maximum $F$. The transmitted, reflected, and total power spectra are shown in red dotted, blue dashed, and black solid curves, respectively. A horizontal line indicating the uncorrelated value $g_2=1$ is plotted. The transmittance at $k=k_\text{peak}$ for the four columns are (from left to right) $37\%, 16.9\%, 67.4\%$, and $54.9\%$. (Parameters used: $\omega_0=\omega_s= 100\Gamma$ and $\Delta=0$.)}
	\label{fig:double 3LS}
\end{figure*}

The last example that we consider is that of two 3LS coupled to an infinite waveguide. A schematic is shown in Fig.\,\figurepanel{fig:schematic}{b},
and the results are in Fig.~\ref{fig:double 3LS}. 

Many of the properties of a two-3LS system are similar to either those of a single 3LS or those of a 3LS plus a mirror. In the first row of Fig.~\ref{fig:double 3LS}, we recognize, for instance, the EIT peak in $T(k)$. $F(k)$ goes to zero at the EIT peak---at perfect transmission there are no correlation effects. We find that the widths of the EIT peaks in both cases are smaller than that of a single 3LS in an infinite waveguide, which qualitatively agrees with the previous studies \cite{FleischhauerRMP05, LeungPRL12}. (Note that in our study the inter-3LS scattering is included.) $T$ and $F$ are symmetric for $k_0L=\pi/2$ ($L=\lambda_0/4$) and distinctly asymmetric for $k_0L=\pi/4$ ($L=\lambda_0/8$). The peak value of $F$ is large, comparable to that for a 3LS plus mirror. The fact that a more complicated structure, providing more interference paths, produces a larger $F_{\rm peak}$, as in these two examples, suggests that it may be a general rule. 

The power spectrum of inelastically scattered radiation upon excitation at $k_{\rm peak}$ is slightly broader than in the other two cases and appears to be formed from overlapping peaks. An important difference with respect to the single 3LS case is that $S(\omega)$ is \emph{not} the same for the reflected and transmitted radiation, though they are of the same order of magnitude. This, then, is the same as was seen for 2LS systems \cite{FangPRA15,LalumierePRA13,vanLooScience13}. In the 2LS case where a photonic band gap builds up as the number of 2LS increases, $S_\text{L}$ (reflection) is always larger than $S_\text{R}$ (transmission) around the 2LS transition frequency. However, Fig.~\figurepanel{fig:double 3LS}{r} shows this is not be true for multiple 3LS.

Turning to the correlation functions (rows 5 and 6), we see that \emph{both} transmitted and reflected photons are bunched at $t=0$ in all cases shown. There is clearly a long time-scale present, in addition to a short one of order $1/\Gamma$. The long time scale is naturally explained in terms of the time delay. Upon inspecting the single-photon transmission coefficient of $N$ identical 3LS (see for example Ref.\,\cite{LeungPRL12}), one finds that the time delay caused by $N$ 3LS is simply $N$ times the single 3LS time delay, $\tau=2N\Gamma/\Omega^2$. Therefore, by increasing the number of 3LS, the time scale of $g_2$ is proportionally lengthened. 

The effect of increasing the classical driving $\Omega$ is very similar to that for a single 3LS: the peaks in $F$ become smaller and shift in frequency, and the correlation functions $g_2$ become more normal with strong bunching or anti-bunching followed by a decay to $g_2 = 1$.

\section{Conclusions}

We have studied non-classical light produced in two 3LS systems---a 3LS in a semi-infinite waveguide and two 3LS in an infinite waveguide---by extending the formalism that was previously applied to multiple 2LS. Sharp peaks in the inelastic flux and the power spectrum appear near the two-photon resonance at which EIT exists. By tuning the classical driving or the qubit-mirror or qubit-qubit separation, these peaks can be made narrower and higher, indicating stronger correlations or ``bound state'' effects. The photon-photon correlation function is studied at the frequency at which the inelastic flux is maximized; it shows complicated bunching and anti-bunching with a very long time-scale. Our study reveals how the slow-light effect associated with EIT is expressed in higher-order quantities such as the correlation function.

Our results show that a 3LS is much more effective than a 2LS in creating photon correlations, making them better candidates for experimental study of the non-classical light produced. Furthermore, we find that the complexity of the structure enhances the photon-photon correlations---they are enhanced by adding additional nonlinear elements (qubits) as well as by simply adding a mirror.

\acknowledgments
This work was supported by U.S.\,NSF Grant No.\,PHY-14-04125.

\appendix
\section{Power spectrum of single 3LS without a mirror}
To check the validity of our scattering approach \cite{ZhengPRL13, FangEPJQT14, FangPRA15}, let us focus on a single 3LS coupled to an infinite waveguide, since it is widely known that the wavefunction of photons propagating in a 1D waveguide coupled to a single, local quantum system can be solved exactly without much difficulty. For this case, we compared our two-photon scattering wavefunction $\langle x_1, x_2|\psi_2\rangle$ with that obtained by directly solving the Schr\"{o}dinger equation in real space \cite{ZhengPRA12}, and we found perfect agreement (not shown).

Furthermore, we find that the power spectrum 
has a simple form when $\Delta=0$ (i.e.\ the classical driving field is not detuned):
\begin{widetext}
\begin{equation}
 \hspace*{-1.45cm}  
S(\omega)=
	\frac{64 (\gamma_++\gamma_-)^4 (E-2 \omega_0)^2 \left[\gamma_+ \gamma_- +8 (\omega-\omega_0) (E-\omega-\omega_0)\right]^2}{\pi ^2 \left(\gamma_+^2+2 (E-2 \omega_0)^2\right) \left(\gamma_-^2+2 (E-2 \omega_0)^2\right) \left(\gamma_+^2+8 (\omega-\omega_0)^2\right) \left(\gamma_-^2+8 (\omega-\omega_0)^2\right) \left(\gamma_+^2+8 (E-\omega-\omega_0)^2\right) \left(\gamma_-^2+8 (E-\omega-\omega_0)^2\right)}
	\label{eq:power spectrum single 3LS}
\end{equation}
with $E=2k_\text{in}$ and 
$\gamma_\pm=\sqrt{\Gamma^2-2\Omega^2\pm\Gamma\sqrt{\Gamma^2-4\Omega^2}}$. We emphasize that this is the power spectrum for \emph{both} left- and right-going photons. 
When $\Omega=0$, this expression for the power spectrum reduces to that of a single 2LS (see Eq.~(A9) of Ref.~\cite{FangPRA15}). In addition, at two-photon resonance ($E=2\omega_0$), $S(\omega)=0$ regardless of $\Omega$.

By integrating over the frequency $\omega$, we then obtain the total inelastically scattered component $F(k_\text{in})$
\begin{equation}
	F(k_\text{in})=2\int S(\omega)d\omega 
	=\frac{16 \sqrt{2} (\gamma_++\gamma_-)^3 (E-2 \omega_0)^2 \left(\gamma_+ \gamma_- +2 (E-2 \omega_0)^2\right)}{\pi  \left(\gamma_+^2+2 (E-2 \omega_0)^2\right)^2 \left(\gamma_-^2+2 (E-2 \omega_0)^2\right)^2}.
	\label{eq:inelastic flux single 3LS}
\end{equation}
It is clear that $F(k)$ decays as $\sim1/k^4$ for $|k_\text{in}-\omega_0|\gg\Gamma$.

These results are used in Fig.~\ref{fig:single 3LS}. We have also derived a general result for $\Delta\neq0$, but it is too lengthy to be reproduced here. Likewise, expressions for $S(\omega)$ and $F(k_\text{in})$ have been obtained for the other two system configurations considered here, but they are also too lengthy.
\end{widetext}

\bibliography{WQED_2015-1108,Dissipation_2015-1108}

\end{document}